\documentclass[12pt]{article}

\usepackage{amsfonts}     
\usepackage{amsmath}
\usepackage{amstext}
\usepackage{amsthm}       %proof environment :)      
\usepackage{xspace}

\input psfig.sty

\newcommand{\N}{\mathbb N}

\newtheorem{teorema}{Theorem}[section]

\theoremstyle{definition}
\newtheorem{definizione}[teorema]{Definition}
\newtheorem{cor}[teorema]{Corollary}

\newtheorem{guess}[teorema]{Remark}

\begin{document}                                            
\title{Computational information for the logistic map at the chaos threshold}
\author{Claudio Bonanno\footnote{Dipartimento di Matematica, Universit\`a 
di Pisa, via Buonarroti 2/a, 56127 Pisa (Italy) - bonanno@mail.dm.unipi.it}  
and Giulia Menconi\footnote{Centro Interdisciplinare per lo Studio dei
Sistemi Complessi, Universit\`a di Pisa, via Bonanno, 25/b  56126 Pisa (Italy)
- menconi@mail.dm.unipi.it}}
\date{}
\maketitle
\begin{abstract}
\noindent We study the logistic map $f(x)=\lambda x(1-x)$ on the unit
square at the chaos threshold. By using the methods of symbolic
dynamics, the information content of an orbit of a dynamical system is
defined as the {\it Algorithmic Information Content (AIC)} of a
symbolic sequence. We give results for the behaviour of the AIC for
the logistic map. Since the AIC is not a computable function we use,
as approximation of the AIC, a notion of information content given by
the length of the string after it has been compressed by a compression
algorithm, and in particular we introduce a new compression algorithm
called CASToRe. The information content is then used to
characterise the chaotic behaviour.
\end{abstract}

\section{Introduction} \label{sI}
Since the discovery of impredictability in deterministic systems that
led to the study of chaotic dynamical systems, much work has been done
to find properties of chaos that could give a classification of these
systems. In particular indicators like Lyapounov exponent,
Kolmogorov-Sinai entropy, topological entropy and others have been
developed.

Nevertheless, in recent years there have been found dynamical systems
for which all the known indicators don't give the presence of chaos,
but numerical results show a high order of impredictability for orbits
of these systems. This phenomenon has led to attempts to generalize the
known indicators (for example see \cite{Tsallis},\cite{Tsallis2}), but
has also stimulated the research for new properties of these systems.

One of the possible new approaches to these {\it weakly chaotic}
dynamical systems is related to information theory, that is to the
notion of information content of a symbolic string. In Section
\ref{sIoss} we give the definition of information content and show how
to apply this technique to dynamical systems.

In particular we study the dynamics of the logistic map (see equation
(\ref{logmap})) at the chaos threshold, that represents one of the best
known examples of weak chaos. In Section \ref{sTlmfipov} we give the
theoretical results that allow a classification of our map among the
weakly chaotic dynamical systems. 

Our theoretical approach is based on the notion of {\it Algorithmic
Information Content} introduced by Kolmogorov and Chaitin (see Section
\ref{sIoss}), and unfortunately this notion of information is not
computable, in the sense that cannot exist an algorithm which computes
this function. Hence to numerically confirm our results, we have to
use an approximate notion of information content. In Section
\ref{sCa}, we present a possible approach to this problem, that
consists on the use of compression algorithms to obtain an estimate
for the information contained in a string. In particular a new
compression algorithm, called CASToRe, is presented and some of its
properties are illustrated. Finally in Section \ref{sErotlm}, we apply
CASToRe to the logistic map at the chaos threshold to confirm the
theoretical predictions for the behaviour of the Algorithmic
Information Content of the strings generated by the map.

\section{Information of symbolic sequences} \label{sIoss}
In our approach the basic notion is the {\it information content of a
symbolic sequence}. Given a finite string $s^n$ of length $n$, that is
a finite sequence of symbols $s_i$, $i=1,\dots,n$, taken in a given
alphabet, the intuitive meaning of {\it quantity of information}
$I(s^n)$ contained in $s^n$ is the following one:
\begin{center}
$I(s^n)$ \textit{is the length of the smallest binary message from which
you can reconstruct} $s^n$.
\end{center}

Formally, this notion leads to the definition of the {\it Algorithmic
Information Content (AIC)} (or {\it Kolmogorov complexity}) introduced
by Chaitin (\cite{Chaitin}) and Kolmogorov (\cite{Kolm}). In order to
define it, it is necessary to define the {\it partial recursive
functions}. We limit ourselves to give an intuitive idea which is very
close to the formal definition. We can consider a partial recursive
function as a computer $C$ which takes a program $p$ (namely a binary
string) as an input, performs some computations and gives a string
$s=C(p)$, written in the given alphabet, as an output. The AIC of a
string $s$ is defined as the shortest binary program $p$ which gives
$s$ as its output, that is
\begin{equation}
I_{AIC}(s,C)=\min\{|p|:C(p)=s\}, \label{eAIC}
\end{equation}
where $|p|$ means the length of the string $p$. A computing machine is
called {\it universal} (namely it is a computer in the usual sense of
the word) if it can simulate any other machine (for a precise
definition see any book on recursion). In particular if $C$ and
$C^{\prime}$ are universal then $I_{AIC}(s,C)\leq
I_{AIC}(s,C^{\prime})+const$, where the constant depends only on $C$
and $C^{\prime}$. This implies that, if $C$ is universal, the
complexity of $s$ with respect to $C$ depends only on $s$ up to a
fixed constant and then its asymptotic behavior does not depend on the
choice of $C$. Thus, we will just write $I_{AIC}(s)$ in equation
(\ref{eAIC}) when referring to the AIC of the string $s$.

We now extend the concept of information to strings generated by a
dynamical system $(X,f)$. Using the usual procedure of symbolic
dynamics, given a partition $\alpha$ of the phase space of the
dynamical system $(X,f)$, it is possible to associate a string
$\Phi_{\alpha}\left( x\right)$ to the orbit having $x \in X$ as
initial condition. If $\alpha=\{A_{1},\dots,A_{l}\},$ then
$\Phi_{\alpha}\left( x\right) =(s_{0},s_{1},\dots,s_{k},\dots)$ if and
only if
\[
f^{k}x\in A_{s_{k}} \quad\forall\ k \ ,
\]
with $s_k\in\{1,\dots l\}\ \forall k$. The AIC of a single orbit of
the dynamical system is then the AIC of the symbolic string generated
by the orbit. If we want not to have dependence on the partition
$\alpha$ of the phase space, then we have to make some procedure of
looking for the supremum varying the partitions. The first results
have been obtained by Brudno (\cite{Brudno}) using open covers of the
phase space. Another possible approach, using computable
partitions\footnote{The intuitive idea of computable partition is a
partition that can be recognised by a computer.}, is introduced in
\cite{Licata}. We are not interested in this problem, but we will just
assume that the partitions we consider are {\it generating} in the
sense that they give the best approximation for the AIC of an orbit of
our dynamical system.

There exist some results connecting the information content of a
string generated by a dynamical system and the Kolmogorov-Sinai
entropy $h^{KS}$ of the system. 

First of all it is proved that in a dynamical system with an ergodic
invariant measure $\mu$ with positive K-S entropy $h^{KS}_\mu$, the
AIC of a string $n$ symbols long behaves like $I_{AIC}(s^n) \sim
h^{KS}_\mu n$ for almost any initial condition with respect to the
measure $\mu$ (\cite{Brudno}).

Instead, in a periodic dynamical system, we expect to find
$I_{AIC}(s^n) = O(\log(n))$. Indeed, the shortest program that
outputs the string $s^n$ would contain only information on the
period of the string and on its length.

It is possible to have also intermediate cases, in which the K-S
entropy is null for all the invariant measures that are physically
relevant and the system is not periodic. These systems, whose
behaviour has been defined {\it weak chaos}, are an important
challenge for research on dynamical systems. Indeed no information are
given by the classical properties, such as K-S entropy or Lyapounov
exponents, and in the last years some generalized definitions of
entropy of a system have been introduced to characterize the behaviour
of such systems (for example see \cite{Tsallis}). We believe that an
approach to weakly chaotic systems using the order of increasing of
their AIC could be a powerful way to classify these systems. This
approach has already been used for the Manneville map (\cite{GW},
\cite{Claudio}):$$f(x)=x+x^z\ (\mbox{mod 1})\qquad x\in [0,1]\ ,\ z\geq 1\ .
$$ In the cited papers it is proved that the AIC of the Manneville map
behaves, for values $z>2$ of its parameter, as
\begin{equation}\label{sporadic} 
I_{AIC}(s^n)\sim n^\alpha, \quad \alpha=\frac{1}{z-1} <1\ ,
\end{equation} 
and this behaviour has been defined {\it sporadicity}.

Among the weakly chaotic systems, one can identify behaviours
different from the sporadic one. As an example, the AIC can have order
smaller than any power law (with respect to the length of the encoded
string). We choose to call this behaviour {\bf mild chaos}.

In this paper we show that the AIC of the logistic map at the chaos
threshold is of order of $\log n$ (Theorem \ref{taic} and Section
\ref{sErotlm}). Then, we will prove that the weakly chaotic dynamics
of the logistic map at the chaos threshold is in particular {\it
mildly chaotic}.

\section{The logistic map from information point of view} \label{sTlmfipov}
We now apply the theory of the Algorithmic Information Content of an
orbit of a dynamical system to the logistic map at the chaos
threshold. We first give some well known results on the dynamics of
the map, and then use these results to obtain an estimate for the AIC.

\subsection{The dynamics of the logistic map} \label{sTdotlm}
The logistic map, defined by 
\begin{equation}
f_\lambda(x)=\lambda x(1-x)\ ,\quad x\in [0,1]\ ,\quad 1 \leq \lambda \leq 4,
\label{logmap}
\end{equation}
is a very simple example of a map with an extremely complicated
behaviour for some values of the parameter $\lambda$. The dynamics of
the map have been studied extensively, and there are many important
results that have been generalized to one dimensional dynamical
systems. Here we give a brief description of the well known period
doubling sequence, and recall some results for the dynamics at the
chaos threshold. The references for the first part of this section are
\cite{Eckmann}, \cite{Demelo}, \cite{Devaney}.

Consider first $\lambda \in (1,3)$. There are two fixed points, $x_0
=0$ and $x_1 = (\lambda -1)/\lambda$. The fixed point at the origin is
unstable and the point $x_1$ is stable. Then every orbit ends up on the
fixed point $x_1$. 

When $\lambda = \lambda_1 = 3$, we have the first bifurcation. The
derivative of $f$ at the point $x_1$ is -1, and the fixed point is now
neutrally stable. Moreover one periodic orbit of period 2 is generated
and it becomes stable as soon as $\lambda$ becomes greater than
$\lambda_1$. This is a period doubling bifurcation.

This kind of bifurcation repeats over and over at different values
$\lambda_n$ of the parameter. That is, when $\lambda \in (\lambda_n,
\lambda_{n+1})$, there is one stable periodic orbit of period $2^{n+1}$, 
and unstable periodic orbits of period $2^j$, $j=0,\dots,n$. When
$\lambda = \lambda_{n+1}$, the stable periodic orbit loses its
stability and a new periodic orbit of period $2^{n+2}$ is generated.

One of the main features of this bifurcation sequence is that the
bifurcation parameters $\lambda_n$ accumulate at $\lambda =
\lambda_{\infty}$, where 
\begin{equation}
\lambda_{\infty} = 3.569945671870944901842\ . \label{lambdainfinito}
\end{equation}
Moreover it holds
\begin{equation} \label{Feigrel}
\lambda_n=\lambda_{\infty}-\frac{b}{\delta ^n},
\end{equation}
where $\delta$ is the so-called {\it Feigenbaum constant} and $b$ is a
suitable constant (\cite{Feig}).

The behaviour of the logistic map at the chaos threshold has attracted
much attention, in particular for being a map with null Kolmogorov
entropy for any invariant probability measure, and nevertheless
showing a weakly chaotic behaviour.

We recall that in this paper by {\it weak chaos} we mean the behaviour
of a map with null Kolmogorov entropy for all the physically relevant
invariant measures and with a dynamics that is neither periodic nor
regular in some sense. As we have seen in Section \ref{sIoss}, a way to
classify weakly chaotic maps is given by the {\it Algorithmic
Information Content (AIC)} of a string generated by the map.

Let's consider now the map $f_{\lambda_\infty}$. For this map there
are countably many unstable periodic orbits of periods $2^j$, for all
$j \in \N$, and an attractor $\Omega$ that is a Cantor set, the
so-called {\it Feigenbaum attractor}. It holds the following theorem
(\cite{Eckmann}, Theorem III.3.5):

\begin{teorema} \label{teckmann}
The logistic map $f_{\lambda_\infty}$ at the chaos threshold has an
invariant Cantor set $\Omega$. 

\noindent (1) There is a decreasing chain of closed subsets 
$$ J^{(0)} \supset J^{(1)} \supset J^{(2)} \supset \dots, $$ each of
which contains $1/2$, and each of which is mapped onto itself by
$f_{\lambda_\infty}$.

\noindent (2) Each $J^{(i)}$ is a disjoint union of $2^i$ closed intervals. 
$J^{(i+1)}$ is constructed by deleting an open subinterval from the
middle of each of the intervals making up $J^{(i)}$.

\noindent (3) $f_{\lambda_\infty}$ maps each of the intervals making 
up $J^{(i)}$ onto another one; the induced action on the set of
intervals is a cyclic permutation of order $2^i$.

\noindent (4) $\Omega = \cap_i J^{(i)}$. $f_{\lambda_\infty}$ maps 
$\Omega$ onto itself in a one-to-one fashion. Every orbit in $\Omega$ is
dense in $\Omega$. 

\noindent (5) For each $k \in \N$, $f_{\lambda_\infty}$ has exactly one 
periodic orbit of period $2^k$. This periodic orbit is repelling and
does not belong to $J^{(k+1)}$. Moreover this periodic orbit belongs
to $J^{(k)} \setminus J^{(k+1)}$, and each point of the orbit belongs
to one of the intervals of $J^{(k)}$.

\noindent (6) Every orbit of $f_{\lambda_\infty}$ either lands after 
a finite number of steps exactly on one of the periodic orbits
enumerated in 5, or converges to the Cantor set $\Omega$ in the sense
that, for each $k$, it is eventually contained in $J^{(k)}$. There are
only countably many orbits of the first type.
\end{teorema}

A characteristic of chaotic dynamical systems is the {\it sensitivity
to initial conditions}. Roughly speaking we can say that a system has
sensitive dependence on initial conditions if two orbits that start
close diverge. 

It is well known that the Kolmogorov-Sinai entropy is related to the
sensitivity to initial conditions of the orbits. The exact relations
between the K-S entropy and the instability of the system is given by
the Ruelle-Pesin theorem. We will recall this theorem in the
one-dimensional case. Suppose that the average rate of separation of
nearby starting orbits is exponential, namely
\[
\Delta x(n)\simeq\Delta x(0)2^{\lambda n}\;\;\;\text{for }\Delta
x(0)\ll 1,
\]
where $\Delta x(n)$ denotes the distance of these two points at time
$n$. If the Lyapounov exponent $\lambda$ is positive then the system
is unstable and $\lambda$ can be considered a measure of its
instability (or sensibility with respect to the initial
conditions). The Ruelle-Pesin theorem implies that, under some
regularity assumptions, $\lambda$ equals the K-S entropy.

The logistic map at the chaos threshold has null Kolmogorov entropy
for any invariant probability measure (it is shown by continuity of
topological entropy and by the Variational Principle), and null
Lyapounov exponent (experimental results), but nevertheless numerical
experiments show that there is a power law divergence for nearby
orbits. This feature of the logistic map has induced the application
of a generalized version of the thermo-dynamical entropy
(\cite{Tsallis}, \cite{Tsallis2}). Formally, we have:

\begin{definizione} \label{sti}
A dynamical system $f:X \to X$ has {\it sensitivity to initial
conditions} if there exists $\delta >0$ such that, for all $x\in X$
and for all neighbourhoods $U$ of $x$, there exist $y \in U$ and $n\in
\N$ such that $d(f^n(x),f^n(y))>\delta$.
\end{definizione}

\begin{teorema} \label{tsti}
The logistic map at the chaos threshold has no sensitivity to initial
conditions. Indeed there exists a subset $X\subset [0,1]$ with
$m(X)=1$, for the Lebesgue measure $m$, such that for all $\delta >0$
and for all $x\in X$, there exists a neighbourhood $U(\delta)$ of $x$
such that $\forall y \in U(\delta)$ and $\forall n\in \N$ we have
$d(f_{\lambda_\infty}^n(x),f_{\lambda_\infty}^n(y))<\delta$.
\end{teorema}

\noindent {\bf Proof.} By Theorem \ref{teckmann}(6), we have that
eventually almost every orbit, with respect to the Lebesgue measure
$m$, ends up on $J^{(k)}$ for all $k$. Moreover by Theorem
\ref{teckmann}(1), we have that $f_{\lambda_\infty}(J^{(k)})\subset
J^{(k)}$.

Let now $\delta >0$ be fixed. Being $\Omega$ a Cantor set, there
exists $k_\delta$ such that $m(J^{(k_\delta)}_i)<\delta$ for all
$i=1,\dots,2^{k_\delta}$, where by $J^{(k_\delta)}_i$ we denote the
intervals making up $J^{(k_\delta)}$. We have (Theorem
\ref{teckmann}(3)) that the action of $f_{\lambda_\infty}$ on
$J^{(k_\delta)}$ is a cyclic permutation of the $2^{k_\delta}$
intervals. Then if $x\in J^{(k_\delta)}_i$, for some $i$, and
$U(\delta)$ is a neighbourhood of $x$ contained in $J^{(k_\delta)}_i$,
then, for all $n$, $f_{\lambda_\infty}^n(U(\delta))\subset
J^{(k_\delta)}_j$, for some $j$. Hence $\forall y \in U(\delta)$ and
$\forall n\in \N$, we have
$d(f_{\lambda_\infty}^n(x),f_{\lambda_\infty}^n(y))<\delta$.

Let now $X$ be the set of the points whose orbit is eventually
contained in $J^{(k_\delta)}$. By Theorem \ref{teckmann}(6), we have
$m(X)=1$. Let now be $x \in X$, then there exists a neighbourhood
$U(\delta)$ of $x$, such that, if $f_{\lambda_\infty}^N(x) \in
J^{(k_\delta)}_i$ for some $i$, then $f_{\lambda_\infty}^N(U(\delta))
\subset J^{(k_\delta)}_i$. The neighbourhood $U(\delta)$ exists by
continuity of the map $f_{\lambda_\infty}$. Now, we can repeat the
same argument as before, showing that $\forall y \in U(\delta)$ and
$\forall n\in \N$, we have
$d(f_{\lambda_\infty}^n(x),f_{\lambda_\infty}^n(y))<\delta$ (if
necessary it is possible to shrink the neighbourhood $U(\delta)$ to
have the previous relation for all $n \in \N$). \qed

\vskip 0.5cm

Hence it remains the problem of explaining the power law divergence of
nearby orbits. One way to give an estimate of the order of this
divergence is by considering the dimension of the set of points that
remain close for some iterations.

\begin{definizione} \label{idc}
For $x \in [0,1]$, we define $$B(x,n,\epsilon) = \{ y \in [0,1] \ | \
d(f_{\lambda_\infty}^i(x),f_{\lambda_\infty}^i(y)) < \epsilon \;
\forall i=0,\dots, n \}.$$
\end{definizione}

Using this definition, we apply Theorem \ref{tsti} to obtain 

\begin{cor} \label{cidc}
For all $\epsilon >0$ and for all $x\in X$, there exists a
$N(x,\epsilon) \in \N$, such that $B(x,n,\epsilon)=B(x,m,\epsilon)$
for all $n,m \ge N(x,\epsilon)$.
\end{cor}

Hence, given a point $x\in X$ and $\epsilon$, the set
$B(x,n,\epsilon)$ shrinks as $n$ increases, until $n$ reaches the
value $N(x,\epsilon)$. At this point the set $B(x,n,\epsilon)$ doesn't
change any more. But, if we are interested in estimate the order of
divergence of nearby orbits at the point $x$, we have to consider the
limit \begin{equation} \label{esti} B(x,n)= \lim_{\epsilon \to 0^+}
B(x,n,\epsilon).\end{equation} We find that $\lim_{\epsilon \to 0^+}
N(x,\epsilon) = +\infty$, for all $x\in X$, hence the function
$B(x,n)$ is a decreasing function of $n$, whose order gives
information on the order of the local divergence. In the next
subsection we derive an estimate for the {\it AIC} of a string
generated by $f_{\lambda_\infty}$, and hence, by a theorem of
\cite{Stefano2}, we find the order of $B(x,n)$.

\subsection{The AIC at the chaos threshold} \label{sTAatct}
In this subsection we show an approach to the dynamics of the logistic
map at the chaos threshold, $f_{\lambda_\infty}$, that allows us to
find an estimate for the AIC (equation (\ref{eAIC})) of almost every
orbit generated by $f_{\lambda_\infty}$. This approach uses the notion
of {\it kneading invariant}. To have a complete treatment of {\it
kneading theory} see, for example, \cite{Demelo}.

Let $f:[0,1]\to [0,1]$ be a $C^1$ {\it unimodal} map, that is there
exists only one point $c\in [0,1]$ such that $f'(c)=0$, and $f$ is
increasing on $(0,c)$ and decreasing on $(c,1)$. We can then define
the {\it kneading sequence} $k_f(x,t)$ of a point $x\in [0,1]$.

\begin{definizione} \label{dks}
For $x\in [0,1]$, we define coefficients $\theta_i(x)$ by: 
$$
\theta_i(x) = 
\left\{
\begin{array}{ll}
+1 & \mbox{ if } \frac{d}{dx} f^{(i+1)} >0 \\[5mm]
-1 & \mbox{ if } \frac{d}{dx} f^{(i+1)} <0 \\[5mm]
0  & \mbox{ if } \frac{d}{dx} f^{(i+1)} =0 \\
\end{array}
\right.
$$
Then the {\it kneading sequence} relative to $x$ is the formal power
series
$$k_f(x,t)= \sum_{i=0}^{\infty} \ \theta_i(x) \ t^i.$$
\end{definizione}

\begin{teorema} \label{tks}
It is possible to introduce a distance $D$ on the space of kneading
sequences, given by $$D(k_f(x,t),k_f(y,t))= \sum_{i=0}^\infty \
\frac{1}{2^i} \ |\theta_i(x) - \theta_i(y)|.$$ This distance makes the
function $x \to k_f(x,t)$ a continuous function for all $x \in [0,1]$,
but the preimages of the critical point $c$.
\end{teorema}

\begin{definizione} \label{dki}
If $x$ is a preimage of the critical point $c$, we define
$$k_f(x_\pm,t)= \lim_{y \to x^\pm} k_f(y,t),$$ where the limit is
taken through sequences of points $y$ that are not preimages of
$c$. For the critical point $c$, it holds
$k_f(c_-,t)=-k_f(c_+,t)$. Then we define the {\it kneading invariant}
$k_f(t)$ of the function $f$ as the sequence $k_f(c_-,t)$.
\end{definizione}

Using classical results on renormalization of maps with null
topological entropy (\cite{Demelo}), one can prove the following:

\begin{teorema} \label{tkiml}
The kneading invariant of the logistic map at the chaos thre\-shold
$f_{\lambda_\infty}$ is given by
\begin{equation} \label{ekiml} 
k_{f_{\lambda_\infty}} (t) = \prod_{i=0}^\infty \ (1-t^{2^i}).
\end{equation}
\end{teorema}

Let's consider now the partition $\alpha =(A_0,A_1)$, given by
$A_0=[0,1/2]$ and $A_1=(1/2,1]$, of the interval $[0,1]$ for the map
$f_{\lambda_\infty}$. If we take a point $x\in [0,1]$, we can code its
orbit into a string $s$ with $s_i \in \{0,1\}$, according to whether
$f_{\lambda_\infty}^i(x)$ is in $A_0$ or in $A_1$.

From the definition of the coefficients $\theta_i(x)$, for any point
$x$, we can see that $\theta_i(x) = \mbox{sgn } (f'(x)\cdot \dots
\cdot f'(f^i(x)))$. Hence, in the coding of the orbit of $x$ into the
string $s$, we have $s_i =0$ if there is a permanence in the sign of
$\theta_{i-1}(x)$ and $\theta_i(x)$, $s_i=1$ otherwise.

We denote by $s(x)$ the string generated by the orbit with initial
condition $x$. Let's start studying the string generated by the point
$x=1/2$, that is the critical point of $f_{\lambda_\infty}$. The
coding of the orbit of $1/2$ into a string $s$ is related to the
kneading invariant of the map $k_{f_{\lambda_\infty}}(t)$, given by
Theorem \ref{tkiml}. Thanks to the particular form of the kneading
invariant (eqn. (\ref{ekiml})), we obtain that to reconstruct the
string $s$, it is sufficient to make the following operations:
\begin{itemize}
\item the first symbol is 0;
\item given $s^n$, that is the first $n$ symbols of the string, we
construct $s^{2n}$ by $s^n \ \bar s^n$, where $\bar s^n$ is equal to
$s^n$ if $n=2^{2k}$, otherwise $\bar s^n$ is obtained by $s^n$
changing the first symbol.
\end{itemize}

Hence we have that $I_{AIC} (s^n(1/2)) = \log n$, indeed it is enough
to specify the length of the string $s^n$. Moreover, by Theorem
\ref{teckmann}(1), we have that $1/2 \in \Omega$, the attractor of
$f_{\lambda_\infty}$, and the orbit of $1/2$ is dense in $\Omega$
(Theorem \ref{teckmann}(4)). Hence for all $x\in X$ (see Theorem
\ref{tsti}), there exists $N(x)$ such that for all $i \ge N(x)$,
$s_i(x)$ can be obtained by $s_i(1/2)$. This implies the following
result:

\begin{teorema} \label{taic}
For almost any point $x\in [0,1]$, the AIC of the string $s(x)$
generated by the orbit of the map $f_{\lambda_\infty}$ with $x$ as
initial condition, is such that 
\begin{equation} \label{eaic}
I_{AIC}(s^n(x))= \log n + const,
\end{equation}
where the constant depends on the point $x$.
\end{teorema}

\begin{guess} \label{raic}
We remark that we have actually proved only the behaviour of the AIC
for the partition $\alpha$. Indeed to obtain the AIC of the orbits of
our dynamical system we should consider the supremum on either all the
open covers or the computable partitions (see Section
\ref{sIoss}). But we have that $\alpha$ is a {\it generating}
partition, that is $\lim_{n\to \infty} diam(\alpha^n) =0$, where
$diam(\alpha)$ is the diameter of a partition, and $\alpha^n = \alpha
\wedge \dots \wedge f^{-n}(\alpha)$. Hence we suppose that the AIC we
estimated is a good approximation of the real one, following the same
arguments used, for example, for the Kolmogorov-Sinai entropy.
\end{guess}

We are now ready to give an estimate on the behaviour of the function
$B(x,n)$ defined in equation (\ref{esti}). Indeed, using Theorem 40 in
\cite{Stefano2}, we have:

\begin{cor} \label{csti}
For almost any point $x\in [0,1]$, the function $B(x,n)$ is of the
order of $n^{-k}$, for some constant $k>0$. Hence, we can say that we
have a power law divergence for nearby orbits, for almost any orbit.
\end{cor}

This result confirms the experiments made on the logistic map at the
chaos threshold. The constant $k$ has been found to be approximately
$1.3236$ (\cite{Tsallis2}).

We have thus derived an estimate for the behaviour of the Algorithmic
Information Content for the logistic map at the chaos threshold,
showing its connection with the sensitivity to initial conditions. We
remark that the AIC is not a computable function, that is it does not
exist an algorithm able to compute the AIC of any string. In the next
section, we show how to obtain an estimate of the AIC for our map.

\section{Compression Algorithms}\label{sCa}
In Section \ref{sIoss} we have introduced the notion of Algorithmic
Information Content (AIC) of a finite string $s$, following the work
of Chaitin (\cite{Chaitin}) and Kolmogorov (\cite{Kolm}). Since the
notion of AIC is not computable, it has been approximated by other
notions of information content of a string that can be computed.

One measure of the information content of a finite string built on an
alphabet ${\cal A}$ can be defined by a lossless (reversible) data
compression algorithm
$$Z:\Sigma({\cal A}) \to \Sigma(\{0,1 \})\ ,$$ that is a coding
procedure such that from the coded string we can reconstruct the
original string. Hence it is natural to consider the length of the
coded string as an approximate measure of the quantity of information
that is contained in the original string. We then define the
information content $I_Z(s)$ of the string $s$ as
\begin{equation} \label{eCIC}
I_{Z}\left(  s\right)  =\left|  Z(s)\right| ,
\end{equation}
where $|Z(s)|$ denotes the length of the compressed string $Z(s)$.
The information content $I_{Z}\left( s\right) $ turns out to be a
computable function and for this reason we will call it Computable
Information Content $(CIC)$. For a more accurate discussion on
compression algorithms and the CIC we refer to \cite{Licata}.

\subsection{A new compression algorithm: CASToRe} \label{sAncac}
In \cite{proceed} and \cite{Licata}, it has been presented a new
compression algorithm, called CASToRe, which has been created in order
to give information on null entropy dynamics and has been used to
study a case of sporadic dynamics, the Manneville map. In the
following, the algorithm CASToRe will be used to evaluate the
Computable Information Content $I_Z$ of the strings (equation
(\ref{eCIC})). As will be better shown later in the case of regular
strings, this new algorithm will be a sensitive measure of the
information content of the string. That's why it is called {\bf
CASToRe}: {\bf C}ompression {\bf A}lgorithm, {\bf S}ensitive {\bf To}
{\bf Re}gularity. In particular, in \cite{Licata} the algorithm
CASToRe has been tested on fully chaotic and sporadic strings,
confirming the theoretical results.

We now give a short description of the algorithm. CASToRe is a
compression algorithm which is a modification of the LZ78 algorithm
(\cite{lz78}).

As it will be proved in Theorem (\ref{teocost}), the Information $I_Z$
of a constant sequence, originally with length $n$, is $4+2\log
(n+1)[\log (\log (n+1))-1]$, if the algorithm $Z$ is CASToRe. The
theory predicts that the best possible information is $I_{AIC}=\log
(n) + $const. In \cite{proceed}, it is shown that the algorithm $LZ78$
encodes a constant $n$ digits long sequence to a string with length
about $const\ +\ n^{\frac 1 2}$ bits; so, we cannot expect that
$LZ78$ is able to distinguish a sequence whose information grows like
$n^{\alpha}$ ($\alpha < \frac 1 2$) (sporadic dynamics) from a
constant or periodic one, while it has been proved (\cite{Licata})
that CASToRe is a very useful tool to identify the correct exponent in
the sporadic case.

The algorithm CASToRe is based on an adaptive dictionary. Roughly
speaking, this means that it translates an input stream of symbols
(the file we want to compress) into an output stream of numbers, and
that it is possible to reconstruct the input stream knowing the
correspondence between output and input symbols. This unique
correspondence between sequences of symbols (words) and numbers is
called \emph{the dictionary}.

At the beginning of encoding procedure, the dictionary is empty. In
order to explain the principle of encoding, let's consider a point
within the encoding process, when the dictionary already contains some
words.

We start analyzing the stream, looking for the longest word W in the
dictionary matching the stream. Then we look for the longest word Y in
the dictionary where W + Y matches the stream. Suppose that we are
compressing an English text, and the stream contains ``basketball
...'', we may have the words ``basket'' (number 119) and ``ball''
(number 12) already in the dictionary, and they would of course match
the stream.

The output from this algorithm is a sequence of word-word pairs (W,
Y), or better their numbers in the dictionary, in our case (119,
12). The resulting word ``basketball'' is then added to the
dictionary, so each time a pair is output to the code-stream, the
string from the dictionary corresponding to W is extended with the
word Y and the resulting string is added to the dictionary.

In the following subsections, we will estimate the information content
$I_Z$, where the algorithm $Z$ is CASToRe, in the case of constant
sequences and periodic sequences with period $p> 1$. Indeed this is
important for the application of CASToRe to the logistic map at the
chaos threshold (Section \ref{sErotlm}).

\subsection{CASToRe on constant strings} \label{sCocs}
We first study the case of period $p=1$ (constant strings). If
we have a string given by
\begin{equation}
s = (aaaaaaaaaaaaaaa.............)\ , \label{successionecostante}
\end{equation}
the algorithm CASToRe gives a codification
\begin{equation}
\begin{array}{llll}
1& :& (0,a) & [a] \\
2& :& (1,1) & [aa] \\
3& :& (2,2) & [aaaa] \\
4& :& (3,3) & [aaaaaaaa] \\
 &  & \dots & \\
k& :& (k-1,k-1) & [a \dots a]
\end{array}
\end{equation} 
so if we compute the information $I_Z(s^n)$, as function of the length
$n$ of the string to be compressed, we can immediately deduce that
$I_Z(s^n) \sim O(\log_2 n)$. Indeed a much more accurate estimate of
the CIC for constant strings is obtained in the following theorem.

\begin{teorema}\label{teocost}
Let $s$ be a constant string as in (\ref{successionecostante}), then
the information function $I_Z(s^n)$ obtained applying the compression
algorithm CASToRe is approximated by the function $\Psi(n)$:
\begin{equation}
\Psi(n) = 4 + 2 \log_2 (n+1) \left( \log_2 (\log_2 (n+1)) -1 \right) , 
\label{infcost}
\end{equation} 
where the approximation is given by the identification of the integer
part of $\log_2 (\log_2 (n))$ with its real value.
\end{teorema}

\noindent {\bf Proof.} We notice that the algorithm CASToRe acts on
$s$ in such a way that at the end of each term of the codification,
the length $n$ of the codified string is of the form $n=2^h -1$. Then
we look for the value of $I(2^h-1)$. We can easily obtain that if
$k=[\log_2 h]$, where $[\cdot]$ denotes the integer part, then
\begin{equation}
\begin{array}{rl}
I(2^h-1)= & 2+ \sum_{i=0}^{k-1} 2 (2^{i+1} -2^i) \log_2 (2^{i+1}) + 2 (n-2^k) \log_2 (2^{k+1}) \\[0.3cm]
= & 2 + \sum_{i=0}^{k-1} (i+1) 2^{i+1} + 2 (n-2^k) \log_2 (2^{k+1}) \\[0.3cm]
= & 4 + (k-1)2^{k+1} + 2 (n-2^k) \log_2 (2^{k+1}),
\end{array} \label{inf1}
\end{equation}
where we used 
\[
\sum_{j=1}^r j 2^j = (r-1) 2^{r+1} +2.
\]
At this point it is enough to approximate $k$ with its real value
$\log_2 h$, and substitute to $h$ its value $\log_2 (n+1)$ to obtain
equation (\ref{infcost}). The theorem is proved. \qed

\subsection{CASToRe on periodic strings} \label{sCops}
We now look at strings $s$ with prime period $p>1$. From the point of
view of the information content, we expect that periodic strings are
not so different from constant strings. Indeed we find that the
algorithm CASToRe is able to recognize periodic strings as constant
strings after having encoded $n_p$ symbols, depending on $p$. Then,
the CIC $I_Z(s^n)$ must behave as in the constant case for $n>n_p$,
indeed it is given by $\Psi(n)$ plus a constant term $C_{s}$ that
depends on the particular string $s$. This term $C_{s}$ can be
estimated as the amount of information obtained from $s$ at the
$n_p$-th symbol. We first prove the following theorem

\begin{teorema}
Let $s$ be a periodic string with prime period $p>1$, then there
exists a constant $n_p$, depending on $s$, such that, when the
algorithm CASToRe reaches the $n_p$-th term of the string, it starts
codifying the string $s$ as if it were a constant string. The
constant $n_p$ has as upper bound
\begin{equation}
K_p=p^2\left [2+\frac{3}{2} \log_2 p + \frac{1}{2}\log_2 |\mathcal{A}|\right ] \label{stimaperiodiche}
\end{equation}
where $|\mathcal{A}|$ is the number of symbols in the alphabet used to
construct the string $s$. Note that $K_p$ depends only on the period
$p$.
\label{teoperiodiche}
\end{teorema}

\noindent {\bf Proof.} Let us assume that $|\mathcal{A}|=r$. Since $s$
has prime period $p$, we can look at it as a sequence of substrings,
each one $p$ symbols long; they will be called $p$-substrings.

Once the codification has started and the first $p$ characters have
been read, the last encoded word will be $t$ symbols long and will end
at some $k$-th (mod. $p$) site in the second $p$-substring. Then,
after at most $(p+1)$ subsequent reading of contiguous $p$-substrings,
the algorithm will arrive at the end of an already known word; this
word will finish at some $h$-th (mod. $p$) site of the $(p+2)$-th
$p$-substring of $s$. 

Let us assume that the number of words needed by the algorithm to
codify the fragment starting from the $h$-th symbol of a previous
$p$-substring to its correspondent belonging to the $(p+2)$-th
$p$-substring of $s$ is $m$; we will indicate the collection of
that $m$ words by $\mathcal{F}(h)$. Let us analyse the different
cases.

{\it First case: $m=2^c$.} The algorithm continues to process the
remaining part of $s$, codifying the next fragment $\mathcal{F}(h)$ by
coupling two words each time. So, after $c$ steps only {\it one} word
will be necessary and later the algorithm will follow the
period.

{\it Second case: $m=2^{c_1} q$, where $q$ is odd}. The algorithm
first behaves $c_1$ times as described above, and after just $q$ words
will be needed to process the following $\mathcal{F}(h)$. Let us call
$A_1,\dots,A_q$ those known words. After one step, the algorithm
CASToRe will have built the new words $B_1=(A_1,A_2),\dots,B_{\frac
{q-1}{2}}=(A_{q-2},A_{q-1})$, and $B_{\frac {q-1}{2}
+1}=(A_q,B_1)$. Then, the algorithm will reach the period after $c_2$
steps, where $c_2 = \min \{ c \in \N : 2^c > q +1 \}$. 

{\it Third case: m is odd.} The algorithm will behave as in the
second case, with $c_1=0$.

So we have proved the existence of the constant $n_p$. Let us estimate
its upper bound, $K_p$. It holds: 
$$n_p\leq p(p+2)[1+C]$$ where $C= \min \{ c \in \N : 2^c > m+1
\}$. Now, we estimate $m$. It does hold $m\leq (p+2)\times S$ where
$S$ is the number of words necessary to cover up a
$p$-substring. Since $r$ different symbols are available in the
alphabet ${\mathcal A}$, then it must be $S\leq N$ where $N$ is the
smaller integer such that it holds $p\leq \sum_{i=1}^N i r^i$.

Since we have $$p \le \frac {r-(N+1)r^{N+1}+Nr^{N+2}}{(r-1)^2} =
\sum_{i=1}^N ir^i $$ and we are interested in an estimation of $N$,
let us look for the smallest integer $N$ such that 
$$\log _r\left( \frac{p(r-1)^2}{r}-1\right )\le 2\ \log_r N .$$
Let $N_0$ the real solution of the following equation:
$$N_0=\sqrt{\frac{(r-1)^2p}{r}-1}\ .$$ 
Then the integer $N$ we looked for is $O(\sqrt{rp})$ and it holds 
$m\leq (p+2)\sqrt{rp}$. So, we have $C\leq 1+\log_2(m+1)$. Finally, 
if we approximate $p+2$ with $p$, we obtain 
$$K_p\leq p^2\left [2 + \frac 3 2 \log_2 p + \frac 1 2 \log_2 r\right ] $$ 
and the theorem is proved. \qed

\section{Experimental results on the logistic map} \label{sErotlm}
In this section we show how to obtain an estimation for the Computable
Information Content $I_Z(s^n)$ for a finite string $s^n$ generated by
the logistic map at the chaos threshold, using CASToRe as compression
algorithm $Z$. In the remainder of this section we drop the subscript
$Z$ from the CIC, since we are always referring to CASToRe.

The plan of our experiments on the logistic map is to apply CASToRe to
strings generated by the logistic map $f_\lambda$ for different values
of $\lambda$ that approximate the chaos threshold $\lambda_\infty$. In
particular we consider two approximating sequences of values of
$\lambda$: the sequence of values $\lambda_j < \lambda_\infty$ at which
the period doubling bifurcations occur (Section \ref{sTpds}) and the
sequence of values $\mu_j > \lambda_\infty$ at which the inverse
tangent bifurcations occur (Section \ref{sItb}). Thus we obtain
approximations for the CIC at the chaos threshold from below and from
above, and the two coincide. We can then say to have obtained an
approximation for the CIC computed with CASToRe for the logistic map
at the chaos threshold $f_{\lambda_\infty}$.

We remark that the symbolic strings are obtained from the orbits of
the logistic map at the chaos threshold considering the partition
$\alpha=(A_0,A_1)$ of $[0,1]$ given by $A_0=[0,1/2)$ and
$A_1=(1/2,1]$. The choice of this partition for the experiments is
justified by its optimality as remarked in Section \ref{sTAatct}.

\subsection{The period doubling sequence}\label{sTpds}
In Section \ref{sCops} we have analysed the behaviour of our algorithm
CASToRe on periodic strings, obtaining that the CIC $I(s)$ of a
periodic string $s$ behaves with respect to the length $n$ of the
encoded string as
\begin{equation} \label{eCICper}
I(s^n) = \Psi(n) + C_s,
\end{equation}
where $\Psi(n)$ is given by equation
(\ref{infcost}) and $C_s$ is a constant depending on the string $s$
(Theorem \ref{teoperiodiche}).

We start considering the logistic map for values of the parameter
$\lambda$ in the sequence $(\lambda_j)_{j\in\N}$ given by the period
doubling bifurcations, for which values the logistic map is
periodic. The sequence $(\lambda_j)_{j\in\N}$ converges to the chaos
threshold value $\lambda_\infty$ from below and can be generated by
the equation (\ref{Feigrel}) (Section \ref{sTdotlm}).

We generated the terms of the sequence $(\lambda_j)_{j\in\N}$ (where
the differences are given up to 38 decimal digits) and iterated the
logistic map $10^7$ times. So we had strings of $10^7$ symbols. When
studying the CIC $I_j(s^n)$ for each $\lambda_j$, where $s^n$ is the
symbolic orbit of the logistic map with parameter $\lambda=\lambda_j$,
we first tried to obtain the constant terms $C_j$, as expected from
equation (\ref{eCICper}). We have obtained an increasing sequence of
values $C_j$ (for $j$ that tends to infinity), and our results about
the compression of periodic strings by the algorithm CASToRe (see
Theorem \ref{teoperiodiche}) are supported by the computations of the
following limits
\begin{equation}
\lim_{n\to \infty} \frac{I_j(s^n) - C_j}
{\Psi(n)} =1 \quad \forall j\in\N \label{limiticoncost}
\end{equation}

As expected, equation (\ref{limiticoncost}) tells us that, after the
compression of a long enough substring of our string $s$, we obtain
the same information function as for constant strings, simply
translated by a constant. If we denote by $n_j$ the number of symbols
that the algorithm has to process before reaching that behaviour, we
have $n_j \to \infty$ as $j \to \infty$. So we have that $C_j \to
\infty$ as $j \to \infty$. 

The approximation of the information content $I_j(s^n)$ with a
function $\Psi(n)$ of the form given by (\ref{infcost}) plus a
constant term is not very good for periodic strings with high
period. This is because of the poorness of the upper bound of our
computations: indeed, we know that this approximation is accurate for
values of $n>n_j$ and there are strings whose $n_j$ is bigger than
$10^7$, the length of our orbits. In our case the period $p_j$ of the
strings $s_j$ is already big enough for $j=12$, indeed $p_{12} =
2^{12} =4098$, so $n_{12}\approx 16\times 10^6$. This means that we
have to look for another way of approximation to the information
function for values of $n$ lower than $n_j$. At the bifurcation points
it is well known from experimental results that the orbit separation
is polynomial and, thanks to the results in \cite{Stefano2}, we can
say that the Algorithmic Information Content is at most
logarithmic. So, we can expect our Computable Information Content to
behave like
\begin{equation}
I_j(s^n) \sim \left\{ 
\begin{array}{ll}
S_j(n) \Psi(n) & \hbox{ for } n<n_j \\[0.3cm]
C_j + \Psi(n) & \hbox{ for } n>n_j
\end{array}
\right.
\quad j \in \N
\label{inftot}
\end{equation}
with $C_j$ as given before. To find the functions $S_j(n)$, we simply
have to compute the fraction
\begin{equation}
S_j(n) = \frac{I_j(s^n)}{\Psi(n)}, \quad j \in \N \ .
\label{Sj}
\end{equation}

\begin{figure}
\begin{tabular}{lr}
{\raggedright{\psfig{figure=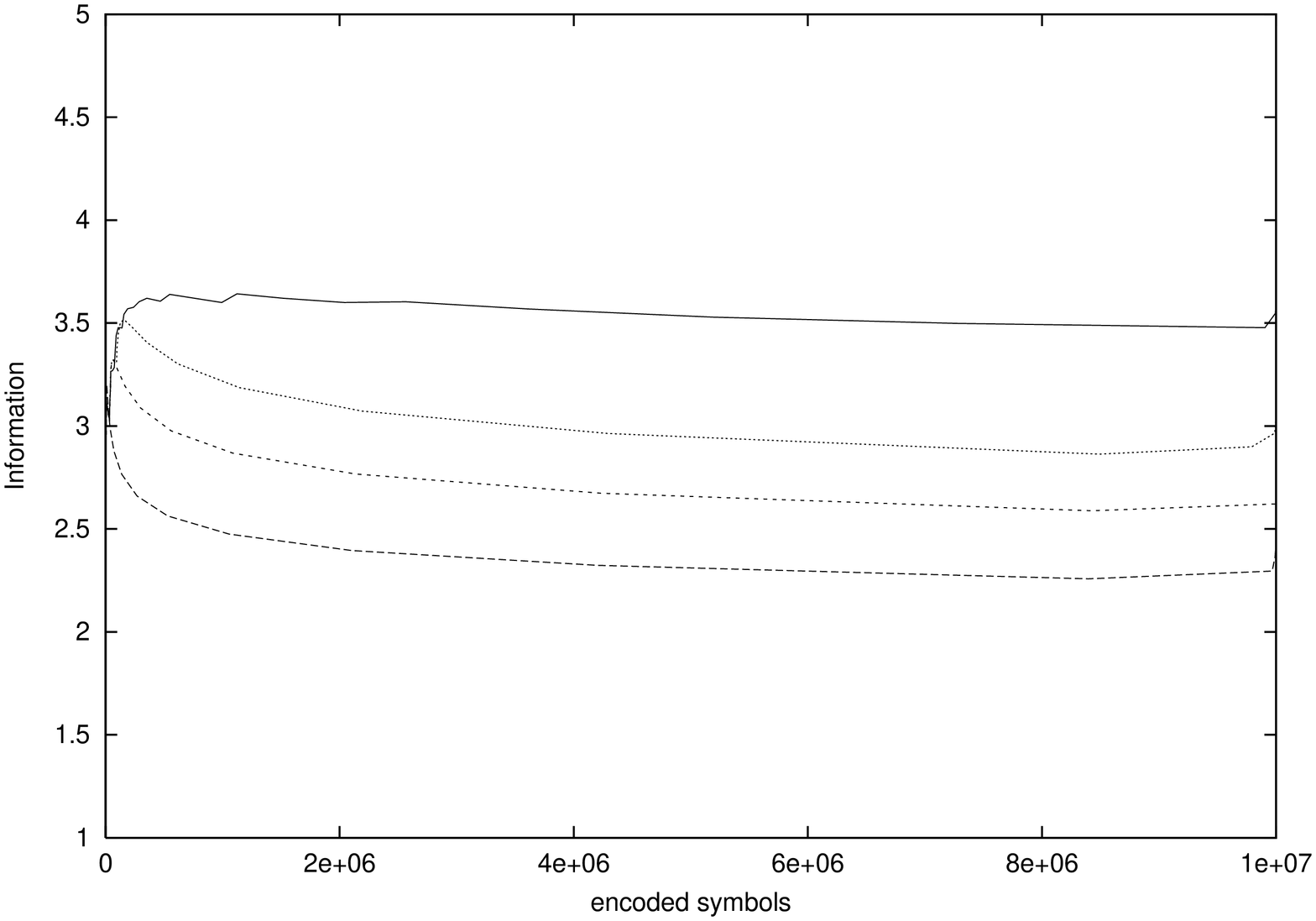,width=6cm,angle=270}}}
&{\raggedleft{\psfig{figure=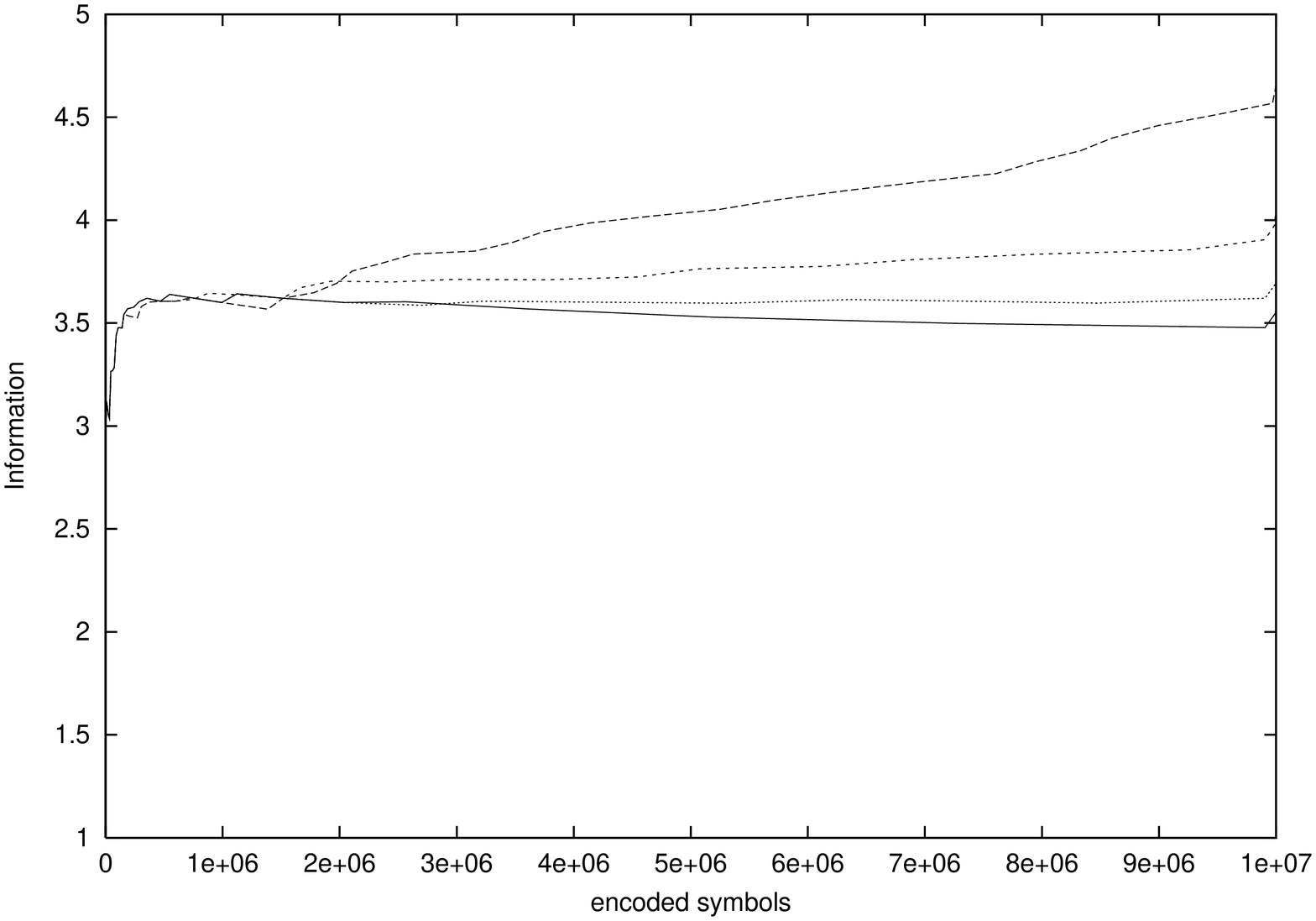,width=6cm,angle=270}}}
\end{tabular}
\caption{\it On the left one can see the behaviour of some of the
functions $S_j (n)$ (dashed lines), that shows the monotonicity of
these functions. The solid line is the function $S_{\infty}(n)$ and
one can see that it is an upper bound for the functions $S_j(n)$. On
the right there is the same picture but for the functions $S_k(n)$
that approximate $S_{\infty}(n)$ from above.}
\label{funzioniS}
\end{figure}

First of all, we deduce from equation (\ref{limiticoncost}) that
$\lim_{n\to \infty} S_j(n)=1$ for all $j$. Moreover the numerical
experiments show that the sequence of functions $(S_j(n))_{j\in\N}$ is
point-wise increasing in $j$ and bounded from above (see figure
\ref{funzioniS} on the left). Then we define the bounding function
$S_\infty (n)$ as
\[
S_\infty (n) = \lim_{j\to \infty} \left( S_j(n) \chi_{_{[0,n_j]}} (n)
+ \chi_{_{[n_j,\infty]}} (n) \right)
\]
for each $n \in \N$, where $\chi_{_{[a,b]}}(n)$ denotes the
characteristic function of the real interval $[a,b]$.

\subsection{Inverse tangent bifurcations}\label{sItb}
In order to establish the behaviour of the information function
$I_{\infty}(s^n)$ at the chaos threshold, we applied the same method as
before, building up a sequence $(\mu_k)_{k\in\N}$ of parameters
approximating $\lambda_{\infty}$ from above.

In the interval $[\lambda_{\infty},4]$, the logistic map has a general
chaotic behaviour, except from narrow ranges (called {\it periodic
windows}) of parameters for which the map is periodic. Inside each
window, a new period doubling sequence can be identified, that leads
the map to chaos. The behaviour becomes periodic from chaotic via a
{\it tangent bifurcation}, that is we can find a fixed point of a
given iterate of the map having 1 as its eigenvalue
(\cite{Pomeau}). The values of the parameter at which the tangent
bifurcations occur converge to $\lambda_{\infty}$ from above and can
be generated using the following relation: 
$$\mu_k=\lambda_{\infty}+\frac{c}{\delta ^k}$$ where $c$ is a suitable
constant and $\delta$ is the Feigenbaum constant.

Using the same definition of $S_k(n)$ given in equation (\ref{Sj}), we
find a sequence of functions such that $\lim_{n\to \infty}
S_k(n)=+\infty$ for all $k\in\N$, due to the chaotic behaviour of the
map at $\mu_k$.

As we can see in figure \ref{funzioniS} on the right, the numerical
evidence is that the sequence $(S_k(n))_{k\in\N}$ is point-wise
decreasing in $k$ and the same function $S_{\infty}(n)$ found in the
previous subsection is a lower bound for the sequence. Then we have
$$\lim_{k\to \infty} S_k(n)=S_{\infty}(n)$$ for each $n$.

\subsection{Results} \label{sR}
From the previous subsections we can draw some conclusions. First of
all, we can say that for all $n \in \N$ the information function
$I_j(s^n)$ of the logistic map (\ref{logmap}) with parameter
$\lambda=\lambda_j$ (as shown in Section \ref{sTpds}) is
\[
I_j(s^n) \sim S_j(n) \Psi(n) \chi_{_{[0,n_j]}} (n) + (\Psi(n) + C_j) \chi_{_{[n_j,\infty]}} (n) \quad \forall \; j \in \N ,
\]
and with parameter $\lambda=\mu_k$ (as shown in
subsection \ref{sItb}) is
\[
I_k(s^n) \sim S_k(n) \Psi(n) \quad \forall \; k \in \N .
\]

Moreover we have that
\[
\lim_{j\to \infty} I_j(n) = \lim_{k\to \infty} I_k(n) = I_\infty (n)= S_\infty (n) \Psi(n) \quad \forall \; n \in \N.
\]

From the numerical approximation of $S_{\infty}(n)$ from above and
below, we conjecture that $S_{\infty}(n)$ is an increasing bounded
function of $n$ and there exists a constant $S_{\infty} =
\lim_{n\to\infty} S_{\infty}(n)$.

\begin{figure}[ht]
\psfig{figure=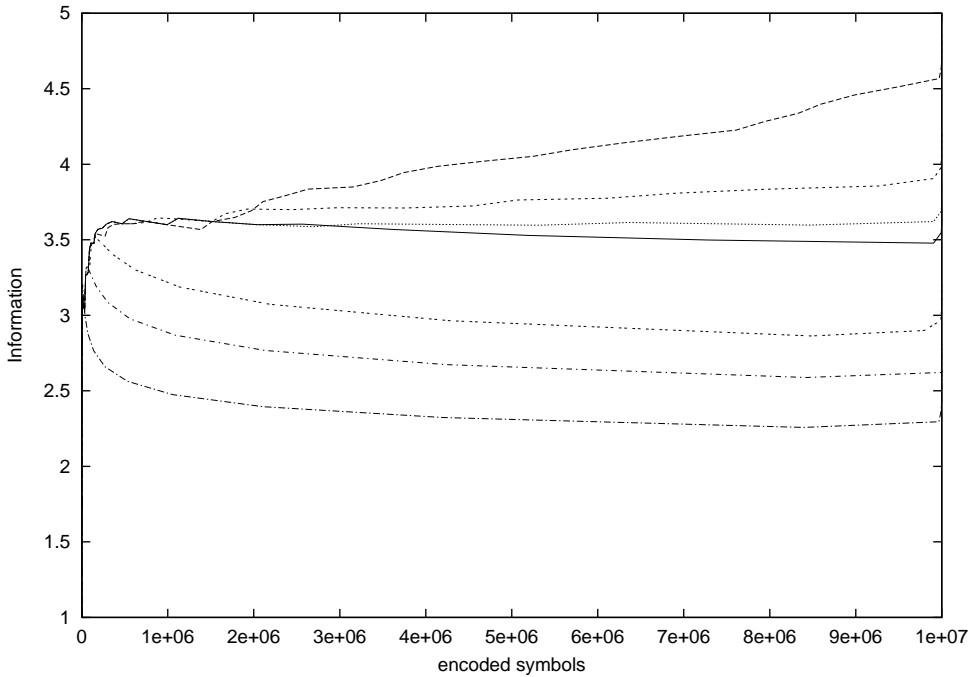,width=13cm,angle=270}
\caption{\it The solid line is the limit function $S_{\infty}(n)$, and
dashed lines are the approximating functions $S_i(n)$ from above and
below.}\label{all}
\end{figure}

Numerical estimates give a value of $S_{\infty}$ of more or less
$3.5$. In figure \ref{all} the behaviour of functions $S_i(n)$
corresponding to some values of both the approximating sequences is
shown.

We have thus confirmed the theoretical result of Theorem \ref{taic},
proving that the logistic map at the chaos threshold is {\it mildly
chaotic} (see Section \ref{sIoss}).

Finally we remark that the feature of $I_{\infty}(s^n)$ and in
particular of $S_{\infty}(n)$ is typical of the logistic map, and is
not simply due to the fact that we are considering periodic orbits
with period going to infinity. Indeed, from the behaviour of the
function $S_{\infty}(n)$, we deduce that $\max_{_n} S_{\infty} (n) <
3.7$ and this means that for $n=10^3$, for example, $I_{\infty}(s^n)$
is less than approximately $148$. But we can construct periodic
strings with period less than $100$, such that the information after
$10^3$ symbols is more than $148$. We can consider, for example, a
string with period given by all the possible combinations of two
symbols in pieces of length from 1 to 4. This string has period $98 <<
2^7$, but $I(734) =356$, much more than $148$.


\begin{thebibliography}{20}

\bibitem{proceed} F. Argenti, V. Benci, P. Cerrai, A. Cordelli,
S. Galatolo, G. Menconi, {\it Information and dynamical systems: a
concrete measurement on sporadic dynamics}, Chaos, Solitons and
Fractals, in press (2001)

\bibitem{Pomeau} P. Berg\'e, Y. Pomeau, C. Vidal, ``Order within
chaos'', Wiley, New York, 1984

\bibitem{Licata} V. Benci, C. Bonanno, S. Galatolo, G. Menconi, F. Ponchio,
{\it Information, complexity and entropy: a new approach to theory and
measurement methods}, http://arXiv.org/abs/math.DS/0107067 (2001)

\bibitem{Claudio} C. Bonanno, {\it The Manneville map: topological, metric 
and algorithmic entropy}, work in preparation

\bibitem{Brudno} A.A. Brudno, {\it Entropy and the complexity of the
trajectories of a dynamical system}, Trans. Moscow Math. Soc.  {\bf 2}
(1983), 127--151

\bibitem{Chaitin} G.J. Chaitin, ``Information, randomness and
incompleteness. Papers on algorithmic information theory'', World
Scientific, Singapore, 1987

\bibitem{Eckmann} P. Collet, J.P. Eckmann, ``Iterated maps on the 
interval as dynamical systems'', Birkh\"auser, 1980

\bibitem{Demelo} W. de Melo, S. van Strien, ``One-dimensional dynamics'',
Springer-Verlag, 1993 

\bibitem{Devaney} R.L. Devaney, ``An introduction to chaotic dynamical
systems'', Addison Wesley, 1987

\bibitem{Feig} M.J. Feigenbaum, {\it Quantitative universality for a class 
of nonlinear transformation}, J. Stat. Phys. {\bf 19} (1978), 25--52

\bibitem{GW} P. Gaspard, X.J. Wang, {\it Sporadicity: Between periodic
and chaotic dynamical behaviors}, Proc. Natl. Acad. Sci. USA {\bf 85}
(1988), 4591--4595

\bibitem{Stefano2} S. Galatolo, {\it Orbit complexity, initial data
sensitivity and weakly chaotic dynamical systems},
http://arXiv.org/abs/math.DS/0102187 (2001)

\bibitem{Kolm} A.N. Kolmogorov, {\it Combinatorial foundations of
information theory and the calculus of probabilities}, Russian
Math. Surveys {\bf 38} (1983), 29--40

\bibitem{lz78} A. Lempel, J. Ziv, {\it Compression of individual sequences
via variable-rate coding}, IEEE Transactions on Information Theory IT
{\bf 24} (1978), 530--536

\bibitem{Tsallis} C. Tsallis, {\it Possible generalization of
Boltzmann-Gibbs statistics}, J. Stat. Phys. {\bf 52} (1988), 479

\bibitem{Tsallis2} C. Tsallis, A.R. Plastino, W.M. Zheng, {\it 
Power-law sensitivity to initial conditions -- new entropic
representation}, Chaos, Solitons, Fractals {\bf 8}, (1987), 885--891

\end{thebibliography}
\end{document}